# The Bretton Woods Experience and ERM

**CHRIS KIRRANE**

**Abstract**

Historical examination of the Bretton Woods system allows comparisons to be made with the current evolution of the EMS. In the past, three main varieties of monetary arrangements have prevailed. One variety is 'decentralised unconstrained systems' in which the sovereignty of each country over its monetary policy is complete (e. g. floating exchange rates arrangements). Another variety is 'fully centralised systems' in which monetary policy is transferred to some supranational authority. Finally, a hybrid variety is 'decentralised but constrained systems' (e. g. the Bretton Woods system and the EMS) where nations are usually (i) sovereign over their domestic policies, but (ii) constrained by a set of widespread rules (for instance, they need to take into account the policies of different nations). This paper attempts to provide a cost-benefit analysis of these various options. Two chief conclusions emerge. Firstly, the apparent autonomy provided} under the 'decentralised unconstrained system' is most likely illusory as the inherent difficulties associated with floating of the various currencies wipe out most of the advantages related to sovereignty. Secondly, 'decentralised but unconstrained systems' such as EMS usually induce a deflationary bias which renders these kinds of arrangements both costly as well as unsustainable. This finally indicates that the fundamental dilemma of global arrangements is that there might not be a stable regime without the construction of a supranational monetary policy institution.

**Introduction**

In the past, international monetary arrangements have involved three types. At one extreme are the systems known as 'decentralised without constraint', where the sovereignty of monetary policies is a totally floating exchange rate regime. At the opposite extreme are fully centralised regimes, where each participant transferred its decision-making to a central authority that defines monetary policy. Between these two extremes, are 'decentralised with constraint' programmes (for example, the EMS and Bretton Woods), where domestic policies are identified individually but must take into consideration the policies of other participants.

This paper appraises these options by discussing their respective advantages and disadvantages. A couple of main conclusions emerge. Firstly, the autonomy of a floating exchange system is probably illusory because floating offsets a substantial part of the benefits of monetary sovereignty. Secondly, decentralised systems with fixed exchange rates are expensive and difficult to sustain.

At the confluence of these two issues is the dilemma fundamental of international systems: a lack of stability without the construction of some kind of supranationality.

Describing the development of the International Monetary Fund, Fred Hirsch [1967] drew the attention of economists to what he referred to as the 'pendulum theory'. In the event that the world is governed by a flexible exchange system, the benefits of fixed exchange rates tend to be put forward by reformers. On the other hand, when you are in a fixed exchange regime, the transition to a floating regime wins votes. Therefore, Bretton Woods succeeded the economic chaos of the 1930s. The floating of the 1970s succeeded Bretton Woods and was replaced by the EMS which dominated Western European trade during the 1980s.

In many respects the proposed monetary union programme proposed by the Maastricht Treaty married again this kind of rhythmic movement of economic thought and policies: at the end of the floatations of the 1970s, there was a gradual convergence of monetary policies in the direction of an irrevocable fixing of exchange rate parities, a prelude to full monetary unification.

At the moment this pendulum effect appears to be temporarily halted. One wonders whether this is a natural slowdown of the motion before the reversal of the tendency, or whether the reversal has already started and the turnaround is already in progress. From all sides voices are raised - in politics, the press, academic circles - to an easing or even suppression of the EMS. The purpose of this article is to take inventory of the oscillations of the pendulum in the long run so as to avoid any short-term approach.

Three main choices seem possible today corresponding to three distinct movements of the pendulum. Firstly, is to go back to a scenario of the 1970s type, in other words a comprehensive decentralised regime. Secondly, would correspond to a temporarily stop in the process of monetary union so as to consolidate achievements without having prejudged the outcome: this is referred to as the decentralised solution with constraint. Thirdly, is to move forward in the direction of a monetary union and to a highly structured regime where the decisions of a supranational authority would replace national policies.

What are the strengths, weaknesses and limitations of these varied options? Two major findings emerge from this analysis which will draw on the historical examples of experiences of Bretton Woods. The first is that to assess the effect of monetary policies and thus their desirability, account should be taken of the political rationale used in their formulation as described below.

The benefits of monetary sovereignty of flotation turn out to be illusory. The second conclusion is that decentralised systems with fixed exchange rates of the EMS type, contain a deflationary bias which makes them expensive or even difficult to sustain. At the confluence of these two problems is situated the fundamental dilemma of international monetary systems; a lack of a well-balanced equilibrium.

**The illusion associated with sovereignty**

The recent troubles of the EMS have led a number of European nations to update their ideas about the advantages of manipulating exchange rates as a monetary policy instrument. Although many economists remain sceptical about the gains - as well as the possibility - of a discretionary monetary policy, a number of experts has claimed to observe the effectiveness of macroeconomic policy and have therefore encouraged Great Britain to hold on to its independence vis-à-vis membership of the EMS (Feldstein [1992]).

This type of advocacy obviously takes various forms but broadly two key approaches can be distinguished, some sort of 'soft' version and a 'radical' version. The soft version assigns to the handling of the exchange rate a passive role of essentially contributing to the process. The radical version, conversely, sets monetary policy an active role of stimulating the economy. The actual distinction, however, is more analytical than truly operational, because it was difficult in practice to fix sufficiently clearly the limit between the two versions: the exchange rate required to regain balance is of course extremely difficult to correctly identify.

Actually, these two analogous versions possess a common origin Mundell (1961) and [1963] and Fleming (1963) analyse economic policy as a means of remedying economic disruptions in different countries. Because these disruptions are heterogeneous and asymmetric, it is necessary that a certain compensatory asymmetry characterises policies to adjust to these shocks. The efficient geography as a function of the analogies between different regions, grouped into 'optimal currency areas' monetary policy as a means of inter-area adjustment and fiscal policy as means of intra-area adjustment. According to the contemporary analysis, the economic heterogeneity of Europe does not make it a currency area, which implies that a plurality of currencies would be preferable to union (see Eichengreen [1990]).

The reappearance of this type of argument in the early 1990s was surprising, since it was precisely scepticism about the effectiveness of autonomous policies that had presided over the creation of the EMS. Moreover, on the margins of intra-SME stability, the macroeconomic effects of floating currencies between Europe and the United States were not predicable. However, exchange rate policymakers' predictions were invalidated by the large appreciation of the dollar after 1981, far from boosting the European economies, it has been accompanied by a slump persistent in Europe, due to a desynchronisation of the cycles between the two continents.

To understand this paradox, it should be stressed that the central assumptions of the theory of the advantages of sovereignty are very weak. Specifically, analyses of the open economy which assign a great deal of power to exchange rate policies assume that companies have no autonomy in terms of prices, and exchange rate fluctuations and their influence on the position of competing candidates. However, a modification of this (See Fitoussi and Le Cacheux [1989]) assumption has a large impact on the conclusion reached. As shown by the Fitoussi and Phelps [1988], the behaviour of corporate margins in imperfect competition is profoundly influenced by the interest rate that sets the terms of the arbitration between present and future profit. Consequently, the higher the real interest rate, the preference for the present is more marked, and therefore companies will try to increase their margin. Similarly, a depreciation, by reducing the pressure of foreign competition, allows national companies to increase their margins without risking losing market share, and therefore future profits. In return, the increase in margins neutralises the effects of the changes in employment. Finally, companies' behaviour to sterilise the stimulus effects associated with exchange rate depreciation, especially when this depreciation is the consequence of a high increase of interest rates abroad.

In this analysis, the traditional Mundell-Fleming model can be reversed. Intuitively, the logic of this result is simple: in a Mundell-Fleming type model, the effect of competitiveness is dominant because firms reflect in terms of production all the increases in demand related to the exchange rate depreciation, which promotes employment. On the contrary, in the model of Phelps and Fitoussi, the effect of depreciation on margins reduces competitiveness, so a flexible exchange rate policy can thus have perverse effects.[1] The importance of the exchange rate regime on the behaviour of aggregate supply and thus employment suggests that the real benefits of exchange rate policy are in fact fairly limited or maybe nonexistent. Analyses of the superiority of discretionary policy norms in line with the work of Barro and Gordon (1983), demonstrate that it is impossible to use a sustainable monetary policy because quite quickly, actors will integrate the discretionary government policy into their forecasts, resulting in the benefits associated with exchange rate policies being neutralised.[2]

This hypothetical argument can be illustrated through empirical observation of the working of the Bretton Woods process: the economic variables (GNP, real interest rates) were significantly more stable compared with other periods (e.g. 1970s), suggesting that that monetary policy may be more efficient, which seems to be paradoxical in the context of a system of fixed exchange rates. Eichengreen [1993] proposed an interpretation of this paradox concerning Fitoussi and Phelps's thesis as to the relationship between the effects of monetary policy and the regime in which it performs. However, this time it was through the influence of expectations - and not on the actions of enterprises - that the demonstration is made.[3] Therefore, monetary policy expansionism in response to a decline in economic activity will have a more favourable effect on employment, when actors do not anticipate that this policy will become systematic. If this is the case, there is no reason for being overly concerned about income erosion due to inflation. This, consequently, limits the degree of effective indexing in-between wages and prices: in such cases the aggregate supply curve represented in the price-production space has a positive slope, so that a little bit of inflation improves competitiveness.

The experience of Bretton Woods thus appears to take the side of supporters of exchange rate policies. According to this particular analysis, the mode conceived in 1944 offered the whole world a nominal anchor and strong stabilisation. Within each European country, the actors perceived inflationary outbursts could not be sustainable as the many various currencies were linked to the dollar, which was itself linked to gold. These expectations allowed stable monetary policies that were not immediately counter-balanced by changes in wages and costs.[4]

---

[1] The influence of the exchange rate regime on the inflation-unemployment trade-off was underlined. For years, the contribution of Dornbusch and Krugman [1976], which showed that in a floating exchange regime, the actual slope of the Phillips curve tended to rise, made arbitration difficult.

[2] Note that the stochastic variants of this theory in which actors guess the government line regarding price and exchange volatility suggest that this kind of argument can be reinforced. The existence of a significant 'noise' surrounding public decisions led to a blurring of perceptions, condemning the government to exaggerate the typical deflations when it wants to make credible its attachment to a virtuous norm - this need to 'shout louder than the crowd' has a substantial social cost.

[3] Note that Eichengreen is actually directly inspired by the article by Phelps [1967] on the interaction between anticipations and arbitrage inflation unemployment.

[4] According to Eichengreen, this analysis is supported by the growing problems of the Capital account, which in the late 1960s were accused of neutralising (through deficits) the consequence of monetary policy. Indeed, the actual reduction of the credibility of the Bretton Woods, making the stability of the nominal anchor far more doubtful, resulted in a tendency to sterilise monetary injections through capital outflows.

Such an interpretation is supported by the study of the persistence of inflation in different monetary regimes, (Alogoskoufis and Smith [1991]). The conclusion is that the persistence of inflation has generally been more problematic in flexible exchange rate regimes than in the typical fixed exchange rate regimes, confirming the notion that in fixed exchange rate regimes, a larger share of inflation is perceived as 'transitory' and thus enhancing the conditions of inflation-unemployment arbitrage. All these analyses thus converge towards a conclusion that the exact form of the aggregate supply function depends on the monetary regime.

In the underlying mechanisms of a system of fixed exchange rates, it seems that the exchange rate regime could deeply affect these mechanisms, tending to counteract the advantages sometimes linked to floating.

It is thus clear that monetary sovereignty, that is, 'Unconstrained' decentralisation does not have large intrinsic advantages: it may not be in itself, but more in comparison with its alternative (which one proposes to call 'Decentralisation with constraint ') that the floating exchange rate system could possibly retain a certain charm. Indeed, as is shown, decentralisation with constraint contains a deflationary bias which has a tendency to propagate the policies of the most of the most restrictive nations in the entire monetary area.

**The deflationary bias associated with decentralised systems with constraint**

One of the chief arguments put forward in order to justify the European central bank, it is one that is quite widespread among European economic experts, (see Casella and Feinstein [1989] or Emerson et al [1992]) which is to claim that decentralisation would be inflationary. Each nation when determining how much money it will issue, will take into account the monetary policies of their competitors. Monetary union subsequently becomes the instrument of an excessive monetary creation, since each member believes that it would have to exploit the passivity of other participants. Indeed, even though profits associated with the monetary issue (the 'seigniorage') come back to the country, the costs (associated with inflation) are shared by the union (see Kirrane [1993]). There would consequently be a problem of coordination which could not be solved other than by the creation of a European central bank.[5]

This ostensibly intuitive argument seems a little artificial and incorrect. Indeed, it results from too much summary analysis of the concepts of decentralisation and also sovereignty. In particular, no nation would agree to go into a multipolar arrangement if it seems to lose any means of control over the decisions of its partners. [6]

---

[5] In a manner much more consistent with reason and practical experience, technically, this is obviously the classic "free rider" problem that arises due to existence of externalities and the possibility of adopting the Nash strategy.

[6] Certainly, a somewhat trivial conclusion of the article by Casella and Feinstein is to say that such an arrangement would be unsound because the participants would prefer flexible exchange rates.

Therefore assume that a decentralised monetary union includes three separate principles:[7] (a) a principle of stability (e.g. Fixed exchange rates between the different members); (B) a principle of sovereignty: each nation is free to make its own policy decisions about monetary policy; And above all (c) a principle of guarantee: each nation is free to determine the amount of foreign currency it is prepared to absorb.[8]

However, as Flandreau has shown [1993], typically the combination of these three elements causes the emergence of a bias that is far from being 'inflationary' and appears rather 'deflationary', in the sense that it imposes a convergence based on the preferences of the most restrictive members' monetary policies. The reason for this is simple: when making a decision, each nation must take into account the impact that it will have on other members, and in return, on itself. Thus, it becomes impossible for a member to engage in excessive monetary creation because it must anticipate that other nations will restrict their absorption of its currency and eliminate it from the system, making use of the guarantee principle. Moreover, each nation may want to exploit the desire of other nations to remain in the monetary union so as to adopt more restrictive policies than they would have wanted to.[9] Such is the deflationary bias of decentralised unions.

History provides so many examples of this thesis that it can be declared that any fixed exchange rate regime has been influenced at any one time by economic deflationary bias. One such example is the Scandinavian Monetary Union (see Kirrane [1993]), established in 1875 between Sweden, Denmark and Norway. In 1885, this union organised the intercirculation of bank notes issued by the different members which were accepted by the central banks of the other member.[10]

Finally, a system of automatically printing banknotes allowed each member nation to finance its deficits through monetary creation: any deficit in payments - for example between Sweden and Norway - gave rise to the creation by one nation's bank, with no obligation to settle immediately. At first glance, the door was wide open to inflationary bias, because nothing seemed to force the debtor nations to adjust their monetary policies. In this arrangement, nevertheless, a principle had been introduced. It consisted of the duty, for each of the Central banks to settle in gold, and on demand of the creditor, the debts it had to other member banks. So the various member nations had to take into account the threat of potentially having to liquidate in a short time in order to redress any imbalance. For example, when Sweden was in debt to Denmark and had to adjust its monetary policies as if the facilities of regulation that had introduced the Scandinavian Union had never existed.

---

[7] Here, "monetary union" will be understood in its broadest meaning to include any fixed exchange rate agreement.

[8] Amount of foreign currency that the monetary bodies are prepared to receive in their accounts, the maximum amount of foreign currency purchased through the foreign exchange market.

[9] The parallelism with this analysis with the traditional debate which makes the EMS an instrument of disinflation is striking as the question raised here is whether the "convergence" of the 1980s was the goal or only the result of the 1979.

[10] A similar proposal was developed in the paper by P. Teles [1993] which implies that the inflationary bias is actually reduced by the introduction of a parallel currency which is used for counteracting agents against the common currency.

In a similar way, the Bretton Woods system illustrates the same mechanism, though in a form perhaps more disguised. The acceptability policy of the dollar standard was ensured by the introduction of a principle of warranty consisting of convertibility of the American currency. Technically, the European nations were protected thus against any temptation by the United States to exploit the system to their profit. In a sense it was legitimate to say (like General de Gaulle had affirmed) that the 'supreme law' of the Bretton Woods system was in effect the key to all policies. It was because of convertibility that the deflationary bias was going to appear. Indeed, it was impossible that the United States engaged in a policy appreciably more inflationist that those of the European nations: the recurrence of deficits making runs on the dollar had fatally put the Europeans in a situation to require payment in gold for their claims. In return, this was to force the Americans to put in place more restrictive deflationary policies or to withdraw from the system altogether. When this occurred, implying a policy change in the United States, and a revaluation of European currencies, Bretton Woods disintegrated. [11] More recently, the last crisis of the European monetary system once again seems to be part of the same analysis. In the framework of the EMS, the principle of warranty is in the rules concerning intervention by the central banks; indeed, the bank of a 'strong' currency is in no case obliged to support without limit, the 'weak' currency. In practice thus, the bank of the weak currency - will always end up devaluing. As, in addition, nothing prevented less inflationary nations from defining their objectives independently, then to impose them on the rest monetary union, while not giving other members the opportunity to leave the system. Thus decentralisation with constraint of fixed exchange rates opens a conflict between national interest and interest of the union, while giving the most restrictive nations a means of solving this conflict 'in their favour' . This is a criticism by supporters of unconstrained decentralisation - but it is not Denmark in the Scandinavian Union, neither France in Bretton Woods, nor Germany in the EMS that is the culprit, it is decentralisation.

Turning to the study of alternative systems to reconcile fixed exchange rates and easing pressures restrictions. Some, in particular, proposed to reintroduce restrictive pressures to improve the 'sustainability' of the EMS. Capital controls would be the real buffer cushion of decentralisation. By modulating their application, one would obtain a range of possible intensities in the transmission of deflationary bias.[12]

At one pole, one would find the 'classical' gold standard, characterised by an absence of controls and resulting in equalisation of interest rates practiced in different nations. Monetary policy decisions are then made very quickly and transmitted from one centre to another through the mediation of the variation in discount rates. If a central bank were to implement a policy to attract gold, the others had to follow the movement. [13] At the other pole, the Bretton Woods System was based on the capital controls which made it possible to achieve compatibility - at least in the short term - on the one hand, monetary or fiscal policy options and, on the other hand, the stability of exchange rates (Giovannini [1989], Giavazzi and Giovannini [1989]). [14]

---

[11] See for example the classical text of Rueff, cited in the article of J. Mr. Jeanneney [1994], and in which Rueff defends the necessary refunding of the gold dollars to save the world from deflation. From 1 January 1951 to December 31st, 1960, the balance of payments deficit of the United States reached 18.1 billion dollars. One could expect that, for this period, the gold reserve decreased by the same amount. However, it rose to 22,8 billion dollars on December 31st, 1950, and it was still, against all probability, at 17.5 billion dollars on December 31st, 1960.

[12] Another factor is the importance of inflation to different nations.

[13] Except to use techniques known as the manipulation of gold points, which can in no case be assimilated to capital controls, but rather to a temporary expansion and limited (less than 1%) of the narrow range where fluctuations change. The integration of European money markets was known as 'Solidarity of the financial markets'. For an analysis of the correlation between the different European interest rates see Flandreau [1994].

These different options are not, however, equivalent in their consequences in terms of well-being, since some impose important distortions. The classic gold standard, although it has occasionally violations, has also enabled emerging economies, such as Japan, Austria or Russia, to name just a few examples, to find necessary financing for their economic development take-off through free access to British, French and German savings. The Bretton Woods system, on the other hand, despite the initial wave of capital transfers, largely has led to a separation of national economies from local savings to finance local investments . [15] Authorities finally find themselves faced with a difficult trade-off between deflation and distortion. Thus, for example, two of the most ardent defenses of capital controls as a means of safeguarding the EMS recognise the difficulty for an economist to advocate throwing sand into the wheels of the international monetary system.[16]

**Necessary Integration**

The necessary integration of the previous analysis allows us to better understand the nature of the oscillations between fixed and floating international arrangements because it shows that none of these balances is, in fact, satisfactory. A decentralised regime with constraint allows the expression of a policy, but not a strategy, because there is no way the mediation of compromise could be reached. At the other pole, a decentralised system without constraint offers the illusion of sovereignty, but without the means of a policy. Also, the instability of changes result more in annihilating than in increasing the effectiveness of monetary policies and increases the need for coordination in order to increase its effectiveness. But as coordination rules 'crystallise', disagreements over the rate of long-term, often short-term inflation, becomes unbearable, and the system collapses: there is no equilibrium natural stability in the functioning of the international monetary system, and every 'regime'" is more or less transitory.

This point of view seems to be confirmed by the observation of Paul De Grauwe [1994],that the credibility of the EMS is not an issue of convergence of inflation rates, but that it is profoundly affected by fluctuations in the unemployment rate in Europe.

---

[14] The oscillations of the pendulum reflect the popularity of movement controls of capital, which fluctuates with that of the advantages of autonomy.

[15] This statement was verified by the mediation of a Feldstein-Horioka test to measure the coefficient of a regression of the rate of saving on the rate investment in various countries. While the coefficient obtained is 1 for the Bretton Woods period, the period 1880-1914 indicates a much lower correlation (about 0.6: see Bayoumi [1990]).

[16] See Eichengreen and Wyplosz, "The Unstable EMS," [1993]. Convergence of inflation rates, reflects the 'pure' decision, nation by nation, to join the EMS. Once taken, this decision imposes on each member a 'discipline' or a 'deflationary bias'"(according to the view taken) in its monetary policy: this is a trade-off between inflation and participation in EMS. On the other hand, fluctuations in employment are a measure of the systemic impact of the deflationary bias at the European level, and the sustainability of the EMS. Not that every government calculates that upon leaving the EMS it could reduce unemployment in a sustainable and sufficient manner but only that the weakening of the credibility of the EMS in the face of employment problems indicates that the nominal anchor emerging from the decentralised structure of Europe is inadequate.

In fact, this inadequacy of decentralisation with constraint is even more evident when the European Monetary System in a broader perspective. Indeed, the logic of adjustment on the most restrictive country is generalised to the context of the response to shocks outside the system. Here again, instead of elaborating a coherent and collective answer, the European nations are reduced to adjust to the policies of less inflationary members: how, in effect, can a country lower its interest rates during the depreciation period of the dollar, if Germany refuses such a policy? The constraint of decentralisation seriously reduces the options available to almost nothing.[17]

Progressively, an analogy between the problems of European unification and the classic dilemmas of the economy can be seen. In essence, the essential difficulty encountered here is no other than that of the implementation of a collective decision-making procedure which meets certain consistency criteria and applies to all. For gold, as Arrow has shown, such a procedure does not exist, unless investing a particular agent the power to decide for others. [18] Obviously the nature of the problems raised by the current operation (or malfunction) of the EMS: failing to have a common policy, the European nations are constrained to align their decisions with those of one of them. This finding, however, has merit, since it indicates the only alternative - the dissolution of the present system and the creation of a European supranationality. [19]

In this regard, the Bretton Woods experiment is very enlightening. Indeed, the debates which surrounded the definition of the statutes of IMF in 1944 illustrated the opposition between Keynes plan (supported by the French delegation: to see Jeanneney [1994]) and the plan Harry Dexter White defended by the United States. While Keynes wanted to equip the IMF with funds of a true supranational power, allowing it to grant appropriations to the nations whose balances of payments would be overdrawn, White's plan was much more restrictive and did not allow such appropriations. Failing to give to the IMF a key role in the management of international liquidity, one thus moved towards a decentralised system on which liquidity would be founded on the dollar.

Only the attempt to build international liquidity on one national currency was fatally to run up against the principle of warranty, because when a national currency is not acceptable it remained convertible into gold. Such is the origin of the false 'problem of liquidities', which developed in the 1960s. Faced with the obviousness of necessary adjustments, the United States rediscovered the charms of the IMF and tried to use it as an instrument in the creation of one international currency. The Special Drawing rights (SDR) were going to be useful substitute to the refunding of the dollars. In the spirit of the Bancor of Keynes, one tried to arrange supranationality in a basically decentralised system. It was understood, it was not a question of a supranationality trompe-l'oeil, or to take again the expression of Marcello de Cecco, as a way 'of eating the cake while claiming that it is still in the refrigerator'.

---

[17] See Fitoussi and Le Cacheux [1989], who develop a model of the economy which posits a 'great country' and a set of 'small countries'.

[18] This refers to the "theorem of the dictator" (Arrow, [1951]). Economists tended to see, in the 1980s, preferences within the EMS as its main advantage, and have praised the he advantage of "tying hands" (see Giavazzi and Pagano [1989], for example). One 'Rediscovers' the structural problem today (long identified by Arrow), which constitutes the attempt to construct a social choice function from preferences of heterogeneous individuals (for an illustration of this type of rediscovery, see Alesina and Grilli [1993]).

[19] It seems superfluous to dwell here on the advantages of the construction of supranationality in the area of monetary negotiations and stability of the international monetary system as they are evident.

What is demonstrated here, it is that there exists in fact probably less options than is believed regarding European integration. One is linked or one is not, and if it is, the union must materialise by a set of precise rules which will be imposed by all (see Kirrane [1994]), that is the construction of supranationality. Any intermediary solution, even if it has chances of functioning perhaps in the medium term, is dedicated early or later to instability.

**Conclusion**

The EMS crisis was a result of its fragile structure. Is it necessary to blame these same structures? Not necessarily. They corresponded to the need for a phased approach, to provide policy with necessary time for convergence. On the other hand, it has been shown that a time of convergence could also become a time of divergence: gradualism is difficult to manage in terms of monetary organisation and it does not last since it is only a medium and not an end. Also, the interpretations of recent difficulties of the EMS that wanted to see it as a sign that the process of integration had to be slowed down have been misunderstood. Thus, while it is recognised that the ultimate goal is unification, it seems necessary to hasten it rather than delay it.